\documentstyle[aps,epsf,twocolumn]{revtex}	 

\begin{document}
\def\inseps#1#2{\def\epsfsize##1##2{#2##1}\centerline{\epsfbox{#1}}}

\twocolumn[\hsize\textwidth\columnwidth\hsize\csname@twocolumnfalse\endcsname

\title{Berry Phase and Ground State Symmetry in $H\otimes h$ Dynamical
Jahn-Teller Systems}

\author{Nicola Manini} \address{European Synchrotron Radiation Facility,
B.P. 220, F-38043 Grenoble C\'edex, France}
\author{Paolo De Los Rios} \address{Institut de Physique Th\'eorique,
Universit\'e de Fribourg, 1700-CH Fribourg, Switzerland}

\date{\today}
\maketitle

\begin{abstract}
Due to the ubiquitous presence of a Berry phase, in most cases of dynamical
Jahn-Teller systems the symmetry of the vibronic ground state is the same
as that of the original degenerate electronic state.  As a single
exception, the linear $H \otimes h$ icosahedral model, relevant to the
physics of C$_{60}$ cations, is determined by an additional free parameter,
which can be continuously tuned to eliminate the Berry phase from the
low-energy closed paths: accordingly, the ground state changes to a
totally-symmetric nondegenerate state.
\end{abstract}

] \narrowtext

The traditional field of degenerate electron-lattice interactions
(Jahn-Teller effect) in molecules and impurity centers in
solids\cite{Englman,Bersuker} has drawn interest in recent years, excited
by the discovery of new systems calling for a revision of a number of
commonly accepted beliefs.  Several molecular systems including C$_{60}$
ions, higher fullerenes and Si clusters, derive their behavior 
from the large (up to fivefold) degeneracy of
electronic and vibrational states due to the rich structure of
the icosahedral symmetry group.\cite{Judd99} Novel Jahn-Teller (JT) systems have
therefore been considered theoretically,\cite{Bersuker,Ihm,AMT} disclosing
intriguing features,\cite{AMT,Mead,Wilczek,Delos96,Moate96} often related
to a Berry phase\cite{Berry} in the electron-phonon coupled dynamics.

As it is well known, the molecular symmetry, reduced by the JT distortion
with the splitting of the electronic-state degeneracy, is restored in 
the dynamical Jahn-Teller (DJT) effect, where
tunneling among equivalent distortions is considered. The vibronic 
states are therefore labelled as
representations of the original point group of the 
undistorted system.  In the weak-coupling regime, for continuity, the
ground state (GS), in particular, retains the same degenerate
representation as that labelling the electronic level prior to coupling.
{\it A priori}, there is no particular reason for this to continue at
larger couplings.  However, it appears empirically\cite{Bersuker} that in
all linear DJT systems studied before the late nineties, the GS symmetry
remains the same at all couplings.  The explanation of this observation was
a great outcome of the Berry-phase\cite{Berry} scenario: the phase
entanglement in the electron-phonon Born-Oppenheimer (BO)
dynamics,\cite{AMT,Ham87,ob92,ob96} originating at 
electronically-degenerate high-symmetry points, seemed a universal feature
of the DJT systems.

In this context, it came as a surprise the discovery of the first linear JT
system showing a {\em nondegenerate GS} in the strong-coupling
limit:\cite{Delos96,Moate96} the spherical model ${\cal D}^{(2)}\otimes
d^{(2)}$ of electrons of angular momentum $L=2$ interacting with vibrations
also belonging to an $l=2$ representation.  This system turns out to be a
special case of the $H\otimes h$ icosahedral model, for a 5-fold degenerate
$H$ electronic state interacting linearly with a distortion mode of the
same symmetry $h$.\cite{Delos96,CeulemansII} In that special case, it was
shown that, for increasing coupling, a nondegenerate $A$ excited state in
the vibronic spectrum moves down, to cross the $H$ GS at some finite value
of the coupling parameter, thus becoming the GS at strong
coupling.\cite{Delos96,Moate96} This phenomenon is a manifestation of the
{\em absence} of Berry-phase entanglement in the coupled
dynamics.\cite{Delos96,noberry}

In this Letter we study the linear $H\otimes h$ model in its generality. We
analyse in detail the connection between the symmetry/degeneracy of the
vibronic GS and the presence/absence of a Berry phase in the coupled
dynamics.  This model owns its peculiarities to the {\em non-simple
reducibility} of the icosahedral symmetry group.  In particular, the $H$
representation appears twice in the symmetric part of the Kronecker product
of the $H$ representation with itself:
\begin{equation}
\lbrace H \otimes H \rbrace^{(s)} = a\oplus g \oplus
			h^{[1]} \oplus  h^{[2]}\;.
\label{kroneker product}
\end{equation}
There are, therefore, two independent sets of Clebsch-Gordan (CG)
coefficients
\begin{equation}
C_{m_1,m_2}^{m~[r]}\; \equiv\; \langle H,m_1;H,m_2|h,m\rangle^{[r]}
\end{equation}
for the coupling of an $H$ electronic state with an $h$
vibrational mode, identified by a multiplicity index
$r=1,2$.\cite{butler81} Of course, since the two $h$ states are totally
equivalent and indistinguishable, symmetry-wise, the choice of these
orthogonal sets of coefficients has some degree of arbitrariness: the free
parameter $\alpha$ in the combination
\begin{equation}
	C_{m_1,m_2}^m\left(\alpha\right)\equiv 
	\cos\alpha ~C_{m_1,m_2}^{m~[1]} + \sin\alpha ~C_{m_1,m_2}^{m~[2]} 
\end{equation}
accounts for it.  The coefficient $C_{m_1,m_2}^m\left(\alpha\right)$
coincides with the $r=1$ and $r=2$ values\cite{butlerconvention:note} for
$\alpha=0$ and $\alpha=\frac \pi 2$ respectively.  Also, for
$\alpha=-\arctan\left( 3 / \sqrt 5 \right)\equiv -\alpha_s$, it becomes
equivalent to the spherical CG coefficient.

The basic Hamiltonian for the $H\otimes h$ model can be written:
\begin{equation}
H = H_{\rm harm}(\hbar \omega) + H_{\rm e-v}(g \hbar \omega,\alpha) \;,
\label{hamiltonian1mode}
\end{equation}
with
\begin{eqnarray}
H_{\rm harm}(\hbar \omega)&=& \frac{1}{2} \hbar \omega 
		\sum_{m} (p_{m}^2 +q_{m}^2) \\
H_{\rm e-v}(g \hbar \omega,\alpha)&=&\frac {g \hbar \omega}2
\sum_{m~m_1 m_2} q_{m} c^\dagger_{m_1} c_{-m_2} 
C_{m_1,m_2}^m\left(\alpha\right) \; ,
\label{interaction hamiltonian}
\end{eqnarray}
where $q_{m}$ is the distortion coordinate (with conjugate momentum $p_{m}$)
and $c^\dagger_{m}$ is the electronic operator in standard second-quantized
notation.

The novelty introduced by the $\alpha$-dependent CG coefficients reflects
the fact that the group does not determine completely the form of the
linear coupling as, for example, in cubic symmetry.  The specific value of
this angle must be established case by case by detailed analysis of the
phonon mode and its coupling with that specific electronic state.  Indeed,
in a realistic case such as, for example, C$_{60}^+$ ions, each $h$ mode is
characterized not only by its own frequency $\omega_i$ and scalar coupling
$g_i$, but also by its particular angle of mixing $\alpha_i$.  

For intermediate to strong coupling, the interesting nonperturbative
regime, the customary framework is the BO separation of vibrational and
electronic motion: when the splitting among the five potential sheets
(proportional to $g^2$) is large, the electronic state can be safely
assumed to follow adiabatically the lowest BO potential sheet, while
virtual inter-sheet electronic excitations may be treated as a small
correction.  The BO dynamics is determined by the lowest eigenvalue of the
interaction matrix $\Xi=\sum q_m V^{(m)}$ in the electronic space. This
matrix is obtained from (\ref{interaction hamiltonian}) by the 
same technique described in Ref.\ \cite{noberry}:
since it is a simple generalization of that obtained for for ${\cal
D}^{(2)}\otimes d^{(2)}$,\cite{noberry} here for brevity we report only the
expression of the diagonal matrix elements of the $V^{(0)}$ matrix,
corresponding to the coupling to a pure $q_0$ distortion:
\begin{equation}
\left[
\matrix{ C_{0,2}^{2}\left(\alpha\right)\cr 
	-C_{0,1}^{1}\left(\alpha\right)\cr
	 C_{0,0}^{0}\left(\alpha\right)\cr
	-C_{0,1}^{1}\left(\alpha\right)\cr
	 C_{0,2}^{2}\left(\alpha\right)\cr  } 
\right]
=
\cos\alpha
\left[
\matrix{ {\frac{1}{2\,{\sqrt{5}}}} \cr {\frac{1}{2\,{\sqrt{5}}}} \cr
{\frac{-2}{{\sqrt{5}}}}\cr
{\frac{1}{2\,{\sqrt{5}}}} \cr 
{\frac{1}{2\,{\sqrt{5}}}} \cr  } 
\right]
+  \sin \alpha
\left[
\matrix{ -{\frac{1}{2}}\cr {\frac{1}{2}} \cr 0 \cr  {\frac{1}{2}} \cr 
-{\frac{1}{2}} \cr  } 
\right] \; .
\label{v0diagonal}
\end{equation}
This form makes it clear that a shift $\alpha\rightarrow \alpha+\pi$
introduces a sign change in the coupling matrix, and it can be compensated
by a reflection $\vec q\rightarrow -\vec q$.  We will restrict therefore,
without loss of generality, to the interval $0\leq \alpha\leq \pi$.

The electronic eigenvalue $\frac{-2}{{\sqrt{5}}} \cos \alpha$ is the lowest
for $\alpha< \alpha_s$ and $\alpha > \pi-\alpha_s$ (region {\bf a}): in
this range the BO potential presents six absolute minima, one of which is
lying along the $\hat q_0$ pentagonal axis, with energy lowering $E_{\rm
clas}=-g^2 / 10 ~ \cos^2 \alpha$ (in units of $\hbar\omega$).
However, influenced by the $V^{(m\neq 0 )}$ matrices, in the complementary
interval $\alpha_s<\alpha<\pi-\alpha_s$ (region {\bf b}), ten trigonal
distortions become the absolute minima, with energy gain $E_{\rm clas}=
-g^2 / 18 ~ \sin^2 \alpha$.  At the boundary angles ($\alpha=\alpha_s$
and $\pi-\alpha_s$), all pentagonal and trigonal minima become degenerate,
and part of a continuous degenerate 4-dimensional (4-D)
trough\cite{Delos96} of depth $E_{\rm clas}= -g^2/28$.

We come now to the r\^ole of the Berry phase in this system. As well known,
the geometrical phase is related to conical degeneracies of the two lowest
BO potential surfaces.\cite{Mead} In the ${\cal D}^{(2)}\otimes d^{(2)}$
system\cite{noberry} (the $\alpha=\pi-\alpha_s$ case of the model studied
here) the flat minimum trough presents {\em tangentially} degenerate
points.  For that case, it was shown that the tangential contacts provide a
mechanism for getting rid of the Berry phase.\cite{noberry,Paris97} For
generic $\alpha$ instead, all contacts between the lowest two potential
sheets occur as conic intersections, instead of tangencies, at points which
are far from the potential minima.  In particular, both trigonal and
pentagonal axes are locations of conical crossings for $\alpha$ in regions
{\bf a} and {\bf b} respectively (i.e.\ when they do not correspond to
minima). In particular, for $\vec q$ on the $\hat q_0$ axis, the five
electronic eigenvalues are given in Eq. (\ref{v0diagonal}), where it can be
readily verified that for $\alpha$ in region {\bf b} the most negative one
is indeed twofold degenerate.

\begin{figure} \epsfxsize 9.0cm \inseps{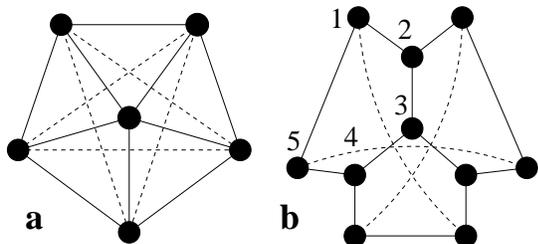}{0.3}
\caption{The connectivity of the BO potential minima for regions {\bf a}
and {\bf b} of angle $\alpha$ introduced in the text.  All lines (solid and
dashed) join nearest neighbor minima.
\label{minima_topology:fig}}
\end{figure}

In region {\bf a}, the six minima are all equidistant, defining the
simplest regular polytope in 5 dimensions (see
Fig.~\ref{minima_topology:fig}a).  In this case, therefore, minimal closed
paths join any of the 20 triplets of minima.  It is straightforward to
verify that at the center of all such triplets there lies one of the
trigonal axes, carrying a conical intersection.  If the degeneracy were
restricted to the trigonal axes, however, the rich topology of the 5-D
space would allow the triangular loop to squeeze continuously to a point
avoiding the degenerate line: the associated Berry phase would then vanish.
Instead, we checked that the two lowest sheets remain in contact through a
bulky 3-D (1 radial + 2 tangential) region of distortions surrounding each
trigonal axis.  This guarantees the nontrivial topology of the loops, thus
the possibility of nonzero Berry phase.
Indeed a {\em geometrical phase of $\pi$} is associated to these triangular
loops, as we computed explicitly by the discretized phase integral of Ref.\
\cite{Resta94}.  Paths encircling two (or any even number) of such triangles
(thus looping through 4, 6,... minima) have zero Berry phase, since the two
phases cancel out.  However, such paths, though energetically equivalent to
the basic triangles (since they cross the same saddle points), are longer,
therefore less relevant from a minimum-action point of view.  We conclude
consequently that, for $\alpha$ in region {\bf a}, the $H\otimes h$ model
must show the signature of a Berry-phase entanglement.

In region {\bf b}, the minima are ten, each with 3 nearest neighbors and 6
second neighbors.  The shortest closed paths through minima joins three
points such as (1$\rightarrow$2$\rightarrow$3$\rightarrow$1) in
Fig.~\ref{minima_topology:fig}b.  However, energetically, such loop is not
the most convenient, since the segment joining two far neighbors
(3$\rightarrow$1) must cross a barrier energetically 60\% more expensive
than that linking next neighbors (1$\rightarrow$2).  Since the energy gaps
between minima and saddle points grow as $g^2$, eventually at strong
coupling only the ``cheapest'' paths affect the low-energy dynamics, and
the relevant Berry phases should be calculated along such loops.  Here,
therefore, at large $g$, the low-energy paths are pentagons, such as
(1$\rightarrow$2$\rightarrow$3$\rightarrow$4$\rightarrow$5$\rightarrow$1)
in Fig.~\ref{minima_topology:fig}b.  We computed the Berry phases for both
kinds of paths, obtaining $\pi$ and $0$ for the 3-points and 5-points loop
respectively.  This implies that the pentagonal loop encircles an even
number (most likely 6) of degenerate regions (one of which around the
pentagonal axis at the center of each 5-points loop), each carrying a phase
factor~$e^{i\pi}$.  We conclude that, in region {\bf b}, although
nontrivial Berry phases are present, they have {\em no effect} on the
strong-coupling low-energy spectrum.  Thus, in particular, the GS symmetry
should remain $H$ in region {\bf a}, while a nondegenerate $A$ state must
turn lower in region {\bf b} at strong coupling.  We stress that we have
established the presence of nonzero Berry phases for all values of
$\alpha$, but also that, in region {\bf b}, the effect of the geometrical
phase is bypassed by energetically cheaper paths with null phase.

\begin{figure} \epsfxsize 9.0cm \inseps{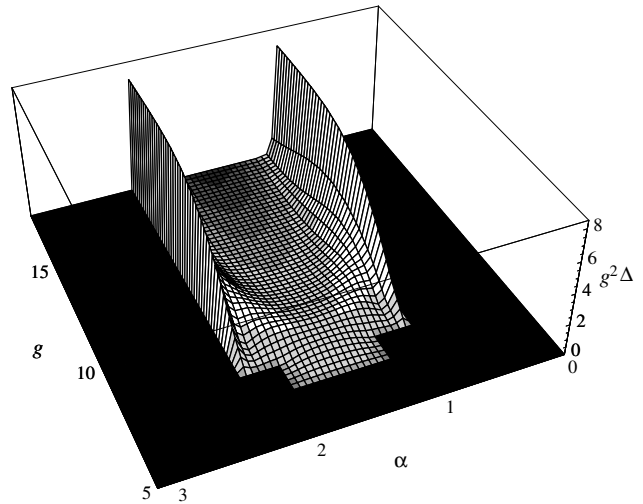}{0.5}
\caption{ $g^2$ times the gap $\Delta=E_H-E_A$ (units of $\hbar\omega$,
logarithmic gray-scale) between the lowest non-$G$ and non-$A$ vibronic
states as a function of $g$ and $\alpha$.  In the positive region, the GS
is $A$, elsewhere it is $H$.  This generates on the $g - \alpha$ plane a
zero-temperature ``phase diagram''.  The basis is truncated to include up
to 40 oscillator states.
\label{1mode40:fig}}
\end{figure}

This scenario is confirmed by numerical diagonalization (Lanczos
method).\cite{Delos96,Reno} In Fig.~\ref{1mode40:fig}, we plot the gap
between the lowest $H$ and $A$ vibronic states, wherever $E_H -E_A>0$, and
0 where the GS is $H$.  At weak coupling, as suggested by continuity, the
GS is $H$.  For $g>7$ and $\alpha$ in range {\bf b}, the $A$ state becomes
the GS.  We note however a little modulation in the boundaries of this
region, both $g$- and $\alpha$-wise.  We observe, in particular, that the
two special values $\alpha_s$ and $\pi-\alpha_s$, far from marking the
closing of the $H-A$ gap, show instead a rather sharp peak in the $\alpha$
direction.  By drawing (in Fig.~\ref{1mode40:fig}) the gap multiplied by
$g^2$, we evidence, along these ridges at $\alpha_s$ and $\pi-\alpha_s$,
the $g^{-2}$ large-$g$ behavior of the $H-A$ gap, characteristic of the
motion in a flat trough of size $\sim g$.  Inside the region {\bf b},
instead, the gap vanishes much more quickly, due to the tunnelling integral
through the barriers between trigonal minima vanishing exponentially
in~$g^2$.
%

It is straightforward to extend the one-mode Hamiltonian
(\ref{hamiltonian1mode}) to a more realistic case of many distortion
modes,\cite{ManyModes} each characterized by its own frequency, coupling
and angle of mixing:
\begin{equation}
H = \sum_{i} \left[ H_{\rm harm}(\hbar \omega_i) + 
		H_{\rm e-v}(g_i \hbar \omega_i,\alpha_i) \right] \; .
\label{hamiltonianmm}
\end{equation}
We study in detail the two-modes case.  Five free parameters ($\omega_1$
being taken as a global scale factor) appear in the model.  In order to
carry out a significant study of the phase diagram, we limit ourselves to
(i) two values only (1 and 5) of the ratio
$\omega_2/\omega_1$,\cite{reflection:note} and (ii)
$\alpha_2-\pi/2=\alpha_1\equiv\alpha$, assuming a principle of "maximum
difference" between the modes.  We take advantage of spectral invariance
for individual sign change of each of the couplings $g_i\rightarrow -g_i$
and for $\alpha \rightarrow -\alpha$, restricting to the
$0\leq\alpha\leq\pi/2$, $g_i>0$ sector.  For convenience, we introduce
polar variables $g_1=g \cos \gamma$, $ g_2=g \sin \gamma$
($0<\gamma<\pi/2$), and draw slices of the parameters space for fixed
values of $g$, as $\alpha - \gamma$ planes.

The first interesting observation concerns the case of equal frequencies:
even though Hamiltonian (\ref{hamiltonianmm}) is linear in the coupling
parameters the CG coefficients and the boson operators, the special case
$\omega_1=\omega_2$ {\em cannot be trivially reduced to a one-mode
problem}, by means of a suitable rotation mixing mode 1 and 2.  This is a
consequence of the linear independence of the coupling matrices
$V^{(m)}(\alpha)$ for different values of $\alpha$.

\begin{figure} \epsfxsize 9.0cm \inseps{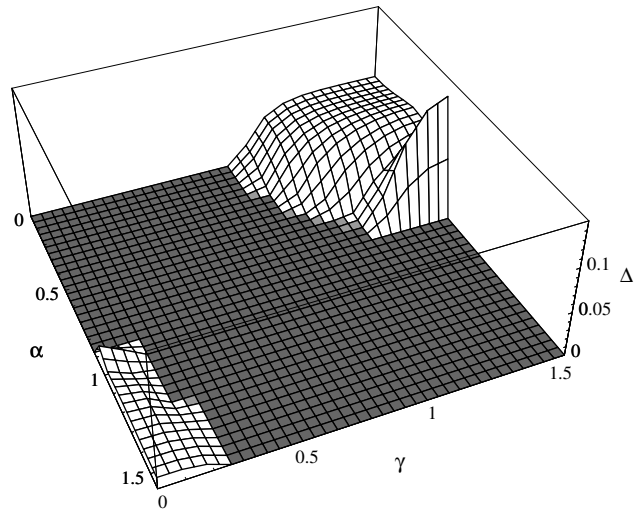}{0.5}
\caption{The gap (units of $\hbar\omega_1$) between the lowest non-$G$ and
non-$A$ vibronic states as a function of $\alpha$ and $\gamma$ (defined in
the text), for $g=10$, $\omega_2/\omega_1 = 5$.  In the positive region,
the GS is $A$, elsewhere it is $H$.  The basis includes up to 12 oscillator
states, enough to give a fairly converged value of $\Delta=E_H-E_A$.
\label{2modes12:fig}}
\end{figure}

We resort to exact diagonalization to treat the two-modes case.  Due to the
larger size of the matrices, we are limited to smaller couplings: we obtain
a satisfactorily converged $E_H -E_A$ gap up to $g\lesssim 10$ only.  The
calculations, for both $\omega_2/\omega_1 = 1$ and 5, show that for $g\leq
7$ the GS symmetry remains $H$ for any $\alpha$ and $\gamma$ as in the
one-mode case.  Then, already at $g=8$, an $A$ (nondegenerate) GS makes its
appearance in two localized regions of the $\alpha-\gamma$ plane.  Starting
from $g\gtrsim 9$, these separated regions assume essentially their
asymptotic strong-coupling shape (see Fig.~\ref{2modes12:fig}).  The first
region, located symmetrically across $\alpha=\pi/2$, corresponds mainly to
mode 1 with {\bf b}-type (no-Berry) coupling:  mode 2 (Berry-phase
entangled in this region) acts as a weak perturbation, incapable to change
the GS symmetry for small enough $\gamma$.  On the other side, the second
region of $A$ GS is located around $\alpha=0$: there, it is mode 2 who is
responsible for the no-Berry phase coupling, mode 1 acting as a weak
perturbation, for $\gamma$ close enough to $\pi/2$.  For
$\omega_2/\omega_1=1$ (not reported here), the two $A$-GS regions are, of
course, equivalent.  For $\omega_2/\omega_1=5$ (Fig.~\ref{2modes12:fig})
instead, these two regions differ in size, in relation with the different
relative energetics of mode 1 versus mode 2.

In conclusion, we have illustrated the importance of the energetics of
paths surrounding the points of degeneracies of the two lowest BO potential
sheets, for defining the effective r\^ole of the Berry phase.  In all
classical linear JT models, the low-energy paths are affected by the
geometrical phase in a way leading to a ``boring'' fixed ground-state
symmetry.  The $H\otimes h$ model is special in being determined by an
additional parameter, allowing to change the connectivity of the graph of
low-energy paths through minima along with the regions of degeneracy of the
two lowest sheets.  Consequently, this new parameter leads continuously
from a regular, Berry-phase entangled, region to a whole region where,
although present, the Berry phase is totally ineffective in imposing its
selection rules to the low-energy vibronic states, and to the GS in
particular.

Finally, for a system such as C$_{60}^+$ our study implies that a detailed
knowledge of not only the coupling parameters $g_i$, but also the
characteristic angles $\alpha_i$ should be acquired for all modes in order
to compute even such a basic property as the GS symmetry.

We thank Arnout Ceulemans, Brian Judd, Fa\-bri\-zia Ne\-gri, Erio To\-sat\-ti,
and Lu Yu for useful discussions.

\bibliographystyle{prsty}

\begin{thebibliography}{10}
\bibitem{butlerconvention:note}
{ We shall stick here to Butler's convention,\cite{butler81} and choose the
  $m_i$ indexes, labelling states within the degenerate representations, by the
  algebraic chain of subgroups $I \supset D5 \supset C5$}.
\end{thebibliography}

\noindent

\noindent

\noindent

\end{document}